\documentstyle[12pt,epsfig]{article}

\catcode`@=11
\def\citer{\@ifnextchar [{\@tempswatrue\@citexr}{\@tempswafalse\@citexr[]}}
 
\def\@citexr[#1]#2{\if@filesw\immediate\write\@auxout{\string\citation{#2}}\fi
  \def\@citea{}\@cite{\@for\@citeb:=#2\do
    {\@citea\def\@citea{--\penalty\@m}\@ifundefined
       {b@\@citeb}{{\bf ?}\@warning
       {Citation `\@citeb' on page \thepage \space undefined}}%
\hbox{\csname b@\@citeb\endcsname}}}{#1}}
\catcode`@=12

\def\refeq#1{\mbox{eq.~(\ref{#1})}}

\def\reffi#1{\mbox{Fig.~\ref{#1}}}

\def\citere#1{\mbox{Ref.~\cite{#1}}}

\newcommand{\mst}{m_{\tilde{t}}}
\newcommand{\delmst}{\Delta\mst}

\newcommand{\mste}{m_{\tilde{t}_1}}
\newcommand{\mstz}{m_{\tilde{t}_2}}

\newcommand{\MstL}{M_{\tilde{t}_L}}
\newcommand{\MstR}{M_{\tilde{t}_R}}

\newcommand{\Mtlr}{M_{t}^{LR}}

\newcommand{\msq}{m_{\tilde{q}}}

\newcommand{\Pe}{\phi_1}
\newcommand{\Pz}{\phi_2}
\newcommand{\PePz}{\phi_1\phi_2}
\newcommand{\mpe}{m_{\Pe}}
\newcommand{\mpz}{m_{\Pz}}
\newcommand{\mpez}{m_{\PePz}}

\newcommand{\oaas}{{\cal O}(\alpha\alpha_s)}
\newcommand{\cp}{{\cal CP}}

\newcommand{\twol}{two-loop}
\newcommand{\onel}{one-loop}

\newcommand{\MW}{M_W}
\newcommand{\MZ}{M_Z}
\newcommand{\MA}{M_A}
\newcommand{\mh}{m_h}

\newcommand{\mt}{m_{t}}
\newcommand{\mtms}{\overline{m}_t}

\newcommand{\mgl}{m_{\tilde{g}}}

\newcommand{\sql}{\tilde{q}_L}
\newcommand{\sqr}{\tilde{q}_R}
\newcommand{\sqe}{\tilde{q}_1}
\newcommand{\sqz}{\tilde{q}_2}

\newcommand{\Stop}{\tilde{t}}

\newcommand{\tst}{\theta_{\tilde{t}}}

\newcommand{\tsf}{\theta\kern-.20em_{\tilde{f}}}
\newcommand{\tsfp}{\theta\kern-.20em_{\tilde{f}\prime}}
\newcommand{\tsq}{\theta\kern-.15em_{\tilde{q}}}

\newcommand{\sinQZtt}{\sin^2 2\tst}

\newcommand{\KL}{\left(}
\newcommand{\KR}{\right)}
\newcommand{\KKL}{\left[}
\newcommand{\KKR}{\right]}

\newcommand{\VL}{\left( \begin{array}{c}}
\newcommand{\VR}{\end{array} \right)}
\newcommand{\ML}{\left( \begin{array}{cc}}
\newcommand{\MLd}{\left( \begin{array}{ccc}}
\newcommand{\MLv}{\left( \begin{array}{cccc}}
\newcommand{\MR}{\end{array} \right)}

\newcommand{\hc}{\mbox {h.c.}}

\newcommand{\Tb}{\tan \beta\hspace{1mm}}

\newcommand{\CTb}{\cot \beta\hspace{1mm}}

\newcommand{\Sb}{\sin \beta\hspace{1mm}}
\newcommand{\SQb}{\sin^2\beta\hspace{1mm}}

\newcommand{\Cb}{\cos \beta\hspace{1mm}}
\newcommand{\CQb}{\cos^2\beta\hspace{1mm}}

\newcommand{\Sa}{\sin \alpha\hspace{1mm}}
\newcommand{\SQa}{\sin^2\alpha\hspace{1mm}}
\newcommand{\Ca}{\cos \alpha\hspace{1mm}}
\newcommand{\CQa}{\cos^2\alpha\hspace{1mm}}

\newcommand{\tev}{\,\, {\mathrm TeV}}
\newcommand{\gev}{\,\, {\mathrm GeV}}

\newcommand{\BC}{\begin{center}}
\newcommand{\EC}{\end{center}}
\newcommand{\BE}{\begin{equation}}
\newcommand{\EE}{\end{equation}}
\newcommand{\BEA}{\begin{eqnarray}}
\newcommand{\BEAnn}{\begin{eqnarray*}}
\newcommand{\EEA}{\end{eqnarray}}
\newcommand{\EEAnn}{\end{eqnarray*}}
\newcommand{\non}{\nonumber}
\newcommand{\id}{{\rm 1\kern-.12em
\rule{0.3pt}{1.5ex}\raisebox{0.0ex}{\rule{0.1em}{0.3pt}}}}

\def\al{\alpha}
\def\als{\alpha_s}

\def\De{\Delta}

\def\hSi{\hat{\Sigma}}

\def\hSie{\hat{\Sigma}^{(1)}}
\def\hSiz{\hat{\Sigma}^{(2)}}

\def\draftdate{\relax}
\def\mda{\relax}
\def\mua{\relax}
\def\mla{\relax}
\def\draft{
\def\thtystars{******************************}
\def\sixtystars{\thtystars\thtystars}
\typeout{}
\typeout{\sixtystars**}
\typeout{* Draft mode!
         For final version remove \protect\draft\space in source file *}
\typeout{\sixtystars**}
\typeout{}
\def\draftdate{\today}
\def\mua{\marginpar[\boldmath\hfil$\uparrow$]%
                   {\boldmath$\uparrow$\hfil}%
                    \typeout{marginpar: $\uparrow$}\ignorespaces}
\def\mda{\marginpar[\boldmath\hfil$\downarrow$]%
                   {\boldmath$\downarrow$\hfil}%
                    \typeout{marginpar: $\downarrow$}\ignorespaces}
\def\mla{\marginpar[\boldmath\hfil$\rightarrow$]%
                   {\boldmath$\leftarrow $\hfil}%
                    \typeout{marginpar: $\leftrightarrow$}\ignorespaces}
\def\Mua{\marginpar[\boldmath\hfil$\Uparrow$]%
                   {\boldmath$\Uparrow$\hfil}%
                    \typeout{marginpar: $\Uparrow$}\ignorespaces}
\def\Mda{\marginpar[\boldmath\hfil$\Downarrow$]%
                   {\boldmath$\Downarrow$\hfil}%
                    \typeout{marginpar: $\Downarrow$}\ignorespaces}
\def\Mla{\marginpar[\boldmath\hfil$\Rightarrow$]%
                   {\boldmath$\Leftarrow $\hfil}%
                    \typeout{marginpar: $\Leftrightarrow$}\ignorespaces}
\overfullrule 5pt
\oddsidemargin -15mm
\marginparwidth 29mm
}

\begin{document}
\thispagestyle{empty}

\def\thefootnote{\fnsymbol{footnote}}

\begin{flushright}
KA--TP--13--1998\\
hep-ph/9807423 \\
%\date{\today}
\end{flushright}

\vspace{1cm}

\begin{center}

{\large\sc {\bf Precise Prediction for the Mass }}

\vspace*{0.4cm} 

{\large\sc {\bf of the Lightest Higgs Boson in the MSSM}}

\vspace{1cm}

{\sc 
S.~Heinemeyer, W.~Hollik and G.~Weiglein} 

\vspace*{1cm}

{\sl
Institut f\"ur Theoretische Physik, Universit\"at Karlsruhe, \\
D--76128 Karlsruhe, Germany}

\end{center}

\vspace*{1cm}

\begin{abstract}
The leading diagrammatic \twol\ corrections are incorporated into the 
prediction for the mass of the lightest Higgs boson, $\mh$, in the Minimal
Supersymmetric Standard Model (MSSM). The results,
containing the complete diagrammatic \onel\ corrections, the new
\twol\ result and refinement terms 
incorporating leading electroweak \twol\ and higher-order QCD
contributions, are discussed 
%in view of the discovery potential of LEP2. They are also 
and compared with results obtained by renormalization
group calculations. Good agreement is found in the case of vanishing
mixing in the scalar quark sector, while sizable deviations occur if 
squark mixing is taken into account.
\end{abstract}
%\pacs{12.15.Lk, 12.60.Jv, 14.80.Cp}

\def\thefootnote{\arabic{footnote}}
\setcounter{page}{0}
\setcounter{footnote}{0}

\newpage

%%%%%%%%%%%%%%%%%%%%%%%%%%%%%%%%%%%%%%%%%%%%%%%%%%%%%%%%%%%%%%
%%%%%%%%%%%%%%%%%%%%%%%%%%%%%%%%%%%%%%%%%%%%%%%%%%%%%%%%%%%%%%

The search for the lightest Higgs boson provides a direct and very 
stringent test of Supersymmetry (SUSY), since the prediction of a relatively
light Higgs boson is common to all Supersymmetric models whose
couplings remain in the perturbative regime up to a very high energy
scale~\cite{susylighthiggs}.
A precise prediction for the mass of the lightest Higgs boson in terms
of the relevant SUSY parameters is crucial in order to determine the
discovery and exclusion potential of LEP2 and the upgraded Tevatron and
also for physics at the LHC, where eventually a high-precision
measurement of the mass of this particle might be possible.

In the Minimal Supersymmetric Standard Model (MSSM)~\cite{mssm} the
mass of the lightest Higgs boson, $\mh$, is restricted at the tree
level to be smaller
than the $Z$-boson mass. This bound, however, is strongly affected by
the inclusion of radiative corrections. The dominant \onel\ corrections
arise form the top and scalar-top sector via
terms of the form $G_F \mt^4 \ln (\mste \mstz/\mt^2)$~\cite{mhiggs1l}. 
They increase the predicted values of
$\mh$ and yield an upper bound of about $150 \gev$. These results have
been improved by performing a complete \onel\ calculation in the
on-shell scheme, which takes into account the contributions of all 
sectors of the MSSM~\cite{mhoneloop,mhiggsf1l,pierce}. 
Beyond \onel\ order renormalization group (RG)
methods have been applied in order to obtain leading logarithmic
higher-order contributions~\cite{mhiggsRG1,mhiggsRG1a,mhiggsRG1b,mhiggsRG2},
and a diagrammatic calculation of the dominant \twol\ contributions in
the limiting case of vanishing $\Stop$-mixing and infinitely large
$\MA$ and $\Tb$ has been carried out~\cite{hoanghempfling}.

The results of the latter calculations were found to yield considerably
lower values for $\mh$ than the \onel\ on-shell calculation. The
results obtained within the RG approach differ by up to $20 \gev$ 
from the \onel\ on-shell result. The bulk of this difference
can be attributed to the higher-order leading logarithmic
contributions which are included in the RG results. The \onel\ result
on the other hand contains non-leading contributions which are not
included in the RG results and whose size has not yet been precisely
determined. Due to the large difference between the complete \onel\
on-shell calculation and the RG results, and due to the difficulty in
comparing the results of the two approaches, it is not easy to give an
estimate for the accuracy of the current theoretical prediction for
$\mh$ in a similar way as done, for instance, for the electroweak
precision observables within the Standard Model (SM) (see \citere{YeRep}). 

Recently a Feynman-diagrammatic calculation of the leading \twol\
corrections of $\oaas$ to the masses of the neutral $\cp$-even Higgs
bosons has been performed~\cite{mhiggsletter}. Compared to the leading
\onel\ result the \twol\ contribution was found to give rise to a 
considerable reduction of the $\mh$ value.

It is the purpose of this letter to combine the full diagrammatic \onel\ 
on-shell result~\cite{mhiggsf1l} with the leading
diagrammatic \twol\ result~\cite{mhiggsletter}, and to obtain in this
way the currently most precise prediction for $\mh$ within the
Feynman-diagrammatic approach for arbitrary values of the parameters of
the Higgs and scalar top sector of the MSSM. Further refinements
concerning the leading \twol\ Yukawa corrections of 
${\cal O}(G_F^2 \mt^6)$~\cite{mhiggsRG1a,ccpw} and of leading QCD
corrections beyond \twol\ order are included. The resulting
predictions for $\mh$ are discussed
in view of the discovery potential of LEP2~\cite{lep2discpot} and 
are compared with the results obtained using the RG 
approach~\cite{mhiggsRG1a,mhiggsRG1b,mhiggsRG2}.
\bigskip

%%%%%%%%%%%%%%%%%%%%%%%%%%%%%%%%%%%%%%%%%%%%%%%%%%%%%%%%%%%%%%
%%%%%%%%%%%%%%%%%%%%%%%%%%%%%%%%%%%%%%%%%%%%%%%%%%%%%%%%%%%%%%

Contrary to the SM, in the MSSM two Higgs doublets are needed.
The  Higgs potential is given by~\cite{hhg}
\BEA
\label{Higgspot}
V &=& m_1^2 H_1\bar{H}_1 + m_2^2 H_2\bar{H}_2 - m_{12}^2 (\epsilon_{ab}
      H_1^aH_2^b + \hc)  \nonumber \\
   && \mbox{} + \frac{g'^2 + g^2}{8}\, (H_1\bar{H}_1 - H_2\bar{H}_2)^2
      +\frac{g^2}{2}\, |H_1\bar{H}_2|^2,
\EEA
where $m_1, m_2, m_{12}$ are soft SUSY-breaking terms, 
$g, g'$ are the $SU(2)$ and $U(1)$ gauge couplings, and 
$\epsilon_{12} = -1$.

The doublet fields $H_1$ and $H_2$ are decomposed  in the following way:
\BEA
H_1 &=& \VL H_1^1 \\ H_1^2 \VR = \VL v_1 + (\phi_1^{0} + i\chi_1^{0})
                                 /\sqrt2 \\ \phi_1^- \VR ,\non \\
H_2 &=& \VL H_2^1 \\ H_2^2 \VR =  \VL \phi_2^+ \\ v_2 + (\phi_2^0 
                                     + i\chi_2^0)/\sqrt2 \VR.
\label{eq:hidoubl}
\EEA
The potential \refeq{Higgspot} can be described with the help of two  
independent parameters (besides $g$, $g'$): 
$\Tb = v_2/v_1$ and $M_A^2 = -m_{12}^2(\Tb+\CTb)$,
where $M_A$ is the mass of the $\cp$-odd $A$ boson.

In order to obtain the $\cp$-even neutral mass eigenstates, the rotation 
\BEA
\VL H^0 \\ h^0 \VR &=& \ML \Ca & \Sa \\ -\Sa & \Ca \MR 
\VL \phi_1^0 \\ \phi_2^0 \VR  
\label{higgsrotation}
\EEA
is performed, where the mixing angle $\alpha$ is given in terms of
$\Tb$ and $M_A$ by
\BE
\tan 2\alpha = \tan 2\beta \frac{\MA^2 + \MZ^2}{\MA^2 - \MZ^2},
\quad - \frac{\pi}{2} < \alpha < 0. 
\EE

%%%%%%%%%%%%%%%%%%%%%%%%%%%%%%%%%%%%%%%%%%%%%%%%%%%%%%%%%%%%%%
%%%%%%%%%%%%%%%%%%%%%%%%%%%%%%%%%%%%%%%%%%%%%%%%%%%%%%%%%%%%%%

At tree level the mass matrix of the neutral $\cp$-even Higgs bosons
is given in the $\phi_1-\phi_2$ basis 
in terms of $\MZ$ and $\MA$ by
\BEA
M_{\rm Higgs}^{2, {\rm tree}} &=& \ML \mpe^2 & \mpez^2 \\ 
                           \mpez^2 & \mpz^2 \MR \non\\
&=& \ML \MA^2 \SQb + \MZ^2 \CQb & -(\MA^2 + \MZ^2) \Sb \Cb \\
    -(\MA^2 + \MZ^2) \Sb \Cb & \MA^2 \CQb + \MZ^2 \SQb \MR,
\EEA
which then has to be rotated with the angle
$\al$ according to \refeq{higgsrotation}, and one obtains the tree-level
Higgs-boson masses
\BE
M_{\rm Higgs}^{2, {\rm tree}} 
   \stackrel{\al}{\longrightarrow}
   \ML m_{H,{\rm tree}}^2 & 0 \\ 0 &  m_{h,{\rm tree}}^2 \MR.
\EE

In the Feynman-diagrammatic approach the one-loop corrected 
Higgs masses are derived by finding the poles of the $h-H$-propagator
matrix whose inverse %(in the $h-H$-basis) 
is given by
\BE
\left(\Delta_{\rm Higgs}\right)^{-1}
= - i \ML q^2 -  m_{H,{\rm tree}}^2 + \hSi_{H}(q^2) &  \hSi_{hH}(q^2) \\
     \hSi_{hH}(q^2) & q^2 -  m_{h,{\rm tree}}^2 + \hSi_{h}(q^2) \MR,
%= \VL \mpe^2 - \hSi_{\Pe}(0)\;\;\;\;\;\; \mpez^2 - \hSi_{\PePz}(0) \\
%     \mpez^2 - \hSi_{\PePz}(0)\;\;\;\;\;\; \mpz^2 - \hSi_{\Pz}(0) \VR ,
\label{higgsmassmatrixnondiag}
\EE
where the $\hSi$ denote the full \onel\ contributions
to the renormalized Higgs-boson self-energies. For these self-energies
we take the result of the complete one-loop on-shell calculation of
\citere{mhiggsf1l}. The agreement with the result obtained in 
\citere{mhoneloop} is better than $1 \gev$ for almost the whole MSSM
parameter space. 

As mentioned above the dominant contribution arises from the
$t-\Stop$-sector. The current eigenstates of the scalar quarks,
$\sql$ and $\sqr$, mix to give the mass eigenstates $\sqe$ and $\sqz$.
The non-diagonal entry in the scalar quark mass matrix is proportional
to the mass of the quark and reads for the $\Stop$-mass matrix
\BE
\mt \Mtlr = \mt (A_t - \mu \CTb), 
\EE
where we have adopted the conventions used in~\citere{drhosuqcd}.
Due to the large value of $\mt$ mixing effects have to be taken into
account. Diagonalizing the $\Stop$-mass matrix one obtains the eigenvalues 
$\mste$ and $\mstz$ and the $\Stop$ mixing angle $\tst$.

The leading \twol\ corrections have been obtained in
\citere{mhiggsletter} by calculating the $\oaas$ contribution 
of the $t-\Stop$-sector to the renormalized Higgs-boson self-energies
at zero external momentum from the Yukawa part of the theory. Since
this calculation was performed in the $\Pe-\Pz$-basis, we perform the
rotation into the $h-H$-basis according to \refeq{higgsrotation}:
\BEA
\hSiz_{H} &=& \CQa \hSiz_{\Pe} + \SQa \hSiz_{\Pz} + 
              2 \Sa \Ca \hSiz_{\PePz} \non \\
\hSiz_{h} &=& \SQa \hSiz_{\Pe} + \CQa \hSiz_{\Pz} - 
              2 \Sa \Ca \hSiz_{\PePz} \non \\
\hSiz_{hH} &=& - \Sa \Ca \KL \hSiz_{\Pe} - \hSiz_{\Pz} \KR + 
              (\CQa - \SQa) \hSiz_{\PePz} .
\label{higgsserotation}
\EEA
%Note that the renormalization of the mixing angle $\al$ does not enter
%the Higgs-mass predictions in $\oaas$.

At the \twol\ level the matrix~(\ref{higgsmassmatrixnondiag}) then
consists of the renormalized Higgs-boson self-energies
\BE
\hSi_s(q^2) = \hSie_s(q^2) + \hSiz_s(0), \quad s = h, H, hH,
\EE
where the momentum dependence is neglected only in the \twol\
contribution.
The Higgs-boson masses at the \twol\ level are obtained by determining the
poles of the matrix $\Delta_{\rm Higgs}$ in \refeq{higgsmassmatrixnondiag}. 

We have implemented two further steps of refinement into the prediction
for $\mh$, which are shown separately in the plots below. 
The leading \twol\ Yukawa correction of ${\cal O}(G_F^2 \mt^6)$ 
is taken over from the result obtained by renormalization
group methods. It reads~\cite{mhiggsRG1a,ccpw}
\BEA
\label{yukawaterm}
\Delta\mh^2 &=& \frac{9}{16\pi^4} G_F^2 \mt^6
               \KKL \tilde{X} t + t^2 \KKR \\
\mbox{with} && \tilde{X} = \Bigg[
                \KL \frac{\mstz^2 - \mste^2}{4 \mt^2} \sinQZtt \KR^2
                \KL 2 - \frac{\mstz^2 + \mste^2}{\mstz^2 - \mste^2}
                      \log\KL \frac{\mstz^2}{\mste^2} \KR \KR \non\\
            && \mbox{}\hspace{1cm}  
               + \frac{\mstz^2 - \mste^2}{2 \mt^2} \sinQZtt
                      \log\KL \frac{\mstz^2}{\mste^2} \KR \Bigg], \\
 && t = \frac{1}{2} \log \KL \frac{\mste^2 \mstz^2}{\mt^4} \KR .
\EEA

The second step of refinement concerns leading QCD corrections beyond
\twol\ order, taken into account by using the $\overline{MS}$
top mass, $\mtms = \mtms(\mt) \approx 166.5 \gev$, for
the \twol\ contributions instead of the pole mass, $\mt = 175 \gev$.
%When performing this refinement the question arises whether $\mtms$
%should be used only directly in the \twol\ result or also indirectly 
%as an input parameter in the $\Stop$ mixing matrix which affects the 
%$\Stop$ masses and the mixing angle in the \twol\ result. Since from
%the viewpoint of an on-shell calculation the latter appears quite
%unnatural, we use this parameterization only in the comparison with the
%RG results below, since in the RG results the running masses appear
%everywhere. In the other plots shown below the refinement of using
%$\mtms$ is understood to be implemented only directly in the \twol\
%result. 
In the $\Stop$ mass matrix, however, we continue to use the pole mass
as an input parameter.
Only when performing the comparison with the RG results we use
$\mtms$ in the $\Stop$ mass matrix for the \twol\ result, since in the
RG results the running masses appear everywhere.
This three-loop effect gives rise to a shift up to $1.5 \gev$ in the
prediction for $\mh$. 

%%%%%%%%%%%%%%%%%%%%%%%%%%%%%%%%%%%%%%%%%%%%%%%%%%%%%%%%%%%%%%%%%%%%%%%
%%%%%%%%%%%%%%%%%%%%%%%%%%%%%%%%%%%%%%%%%%%%%%%%%%%%%%%%%%%%%%%%%%%%%%%

For the numerical evaluation we have chosen two values for $\Tb$ which
are favored by SUSY-GUT scenarios~\cite{su5so10}: $\Tb = 1.6$ for
the $SU(5)$ scenario and $\Tb = 40$ for the $SO(10)$ scenario. Other
parameters are $\MZ = 91.187 \gev, \MW = 80.375 \gev, 
G_F = 1.16639 \, 10^{-5} \gev^{-2}, \als(\mt) = 0.1095$, and $\mt = 175 \gev$. 
For the figures below we have furthermore
chosen $M = 400 \gev$ ($M$ is the soft SUSY breaking parameter in the
chargino and neutralino sector), $\MA = 500 \gev$,
and $\mgl = 500 \gev$ as typical 
values (if not indicated differently). The scalar top masses and the
mixing angle are derived from the 
parameters $M_{{\tilde t}_L}$, $M_{{\tilde t}_R}$ and $\Mtlr$ of the 
$\Stop$ mass matrix (our conventions are the same as in
\citere{drhosuqcd}). In the figures below we have chosen 
$\msq \equiv \MstL = \MstR$.

\begin{figure}[htb]
\begin{center}
\mbox{
\psfig{figure=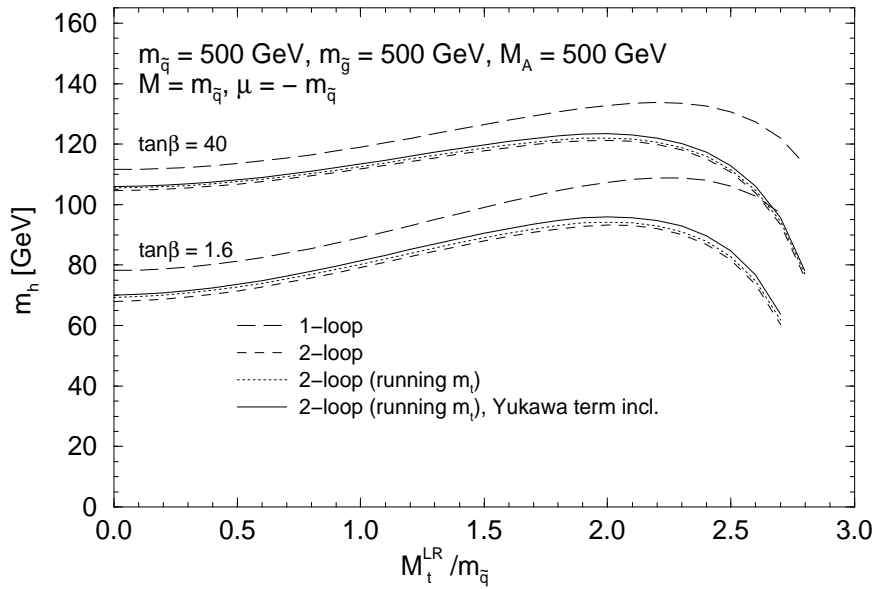,width=7.3cm,height=8cm,
                      bbllx=150pt,bblly=100pt,bburx=450pt,bbury=420pt}}
\end{center}
\caption[]{One- and \twol\ results for $\mh$ as a function of
$\Mtlr/\msq$ for two values of $\Tb$. The two steps of refinement
discussed in the text are shown separately.}
\label{fig:plot1}
\end{figure}

The plot in Fig.~\ref{fig:plot1} shows the result for $\mh$ obtained
from the diagrammatic calculation of the full \onel\ and leading
\twol\ contributions. The two steps of refinement discussed above are
shown in separate curves. For comparison the pure \onel\ result is also
given. The results are plotted as a function of $\Mtlr/\msq$, where
$\msq$ is fixed to $500 \gev$. The qualitative behavior is the same as 
in \citere{mhiggsletter}, where the result containing only the leading
\onel\ contribution (and without further refinements) was shown.
The \twol\ contributions give rise to a large reduction of the 
\onel\ result of 10--20 GeV. The two steps of refinement both
increase $\mh$ by up to $2 \gev$. 
A minimum occurs for  $\Mtlr = 0 \gev$ which we refer to as `no mixing'. 
A maximum in the \twol\ result for $\mh$ is reached for about 
$\Mtlr/\msq \approx 2$ in the $\Tb = 1.6$ scenario as well as in the 
$\Tb = 40$ scenario. This case we refer to as `maximal mixing'. 
The maximum is shifted compared to its \onel\ value of about
$\Mtlr/\msq \approx 2.4$. The two steps of refinement have only a
negligible effect on the location of the maximum. 

%%%%%%%%%%%%%%%%%%%%%%%%%%%%%%%%%%%%%%%%%%%%%%%%%%%%%%%%%%%%%%
%%%%%%%%%%%%%%%%%%%%%%%%%%%%%%%%%%%%%%%%%%%%%%%%%%%%%%%%%%%%%%

\begin{figure}[htb]
\begin{center}
\hspace{1em}
\mbox{
\psfig{figure=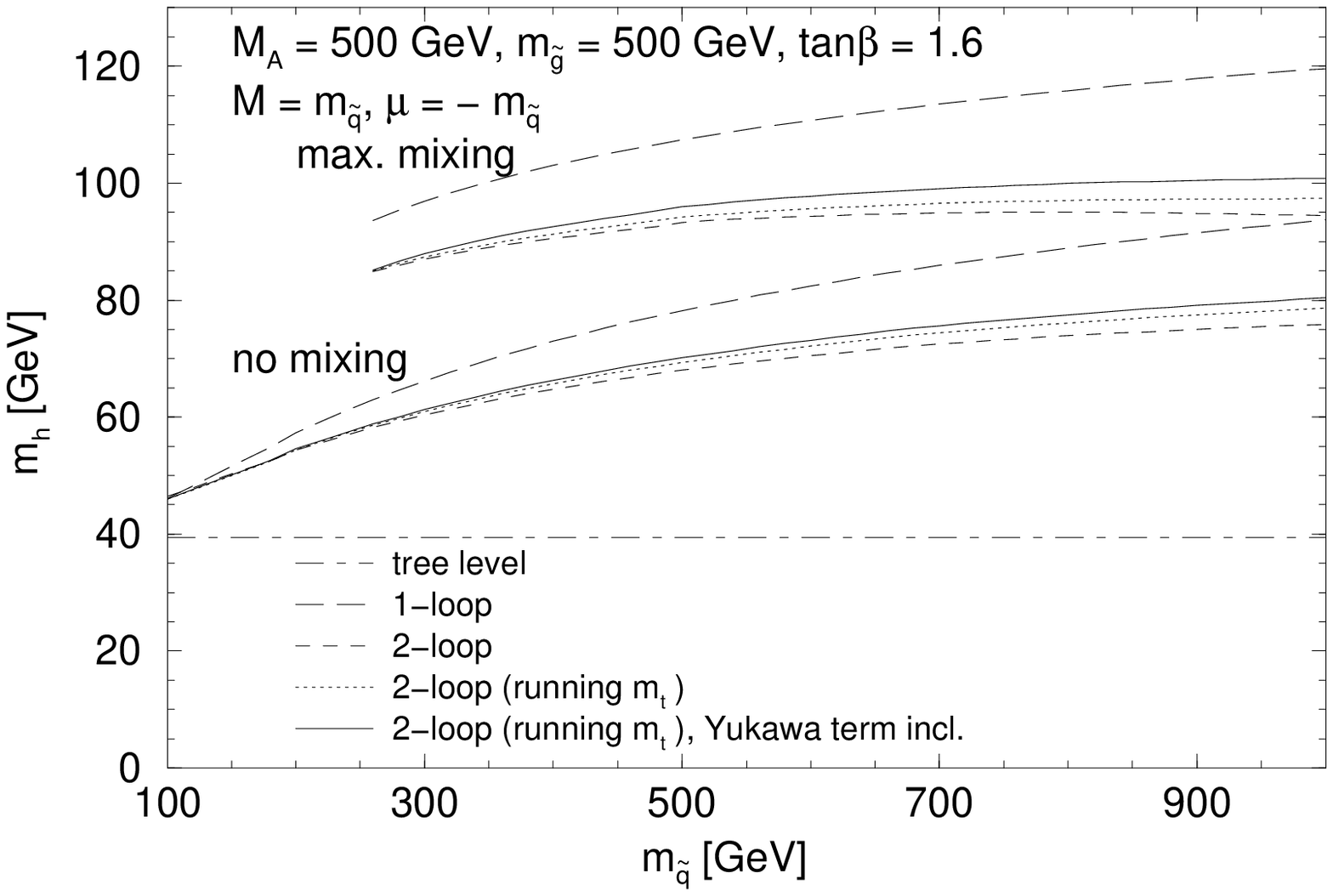,width=5.3cm,height=8cm,
                      bbllx=150pt,bblly=100pt,bburx=450pt,bbury=420pt}}
\hspace{7.5em}
\mbox{
\psfig{figure=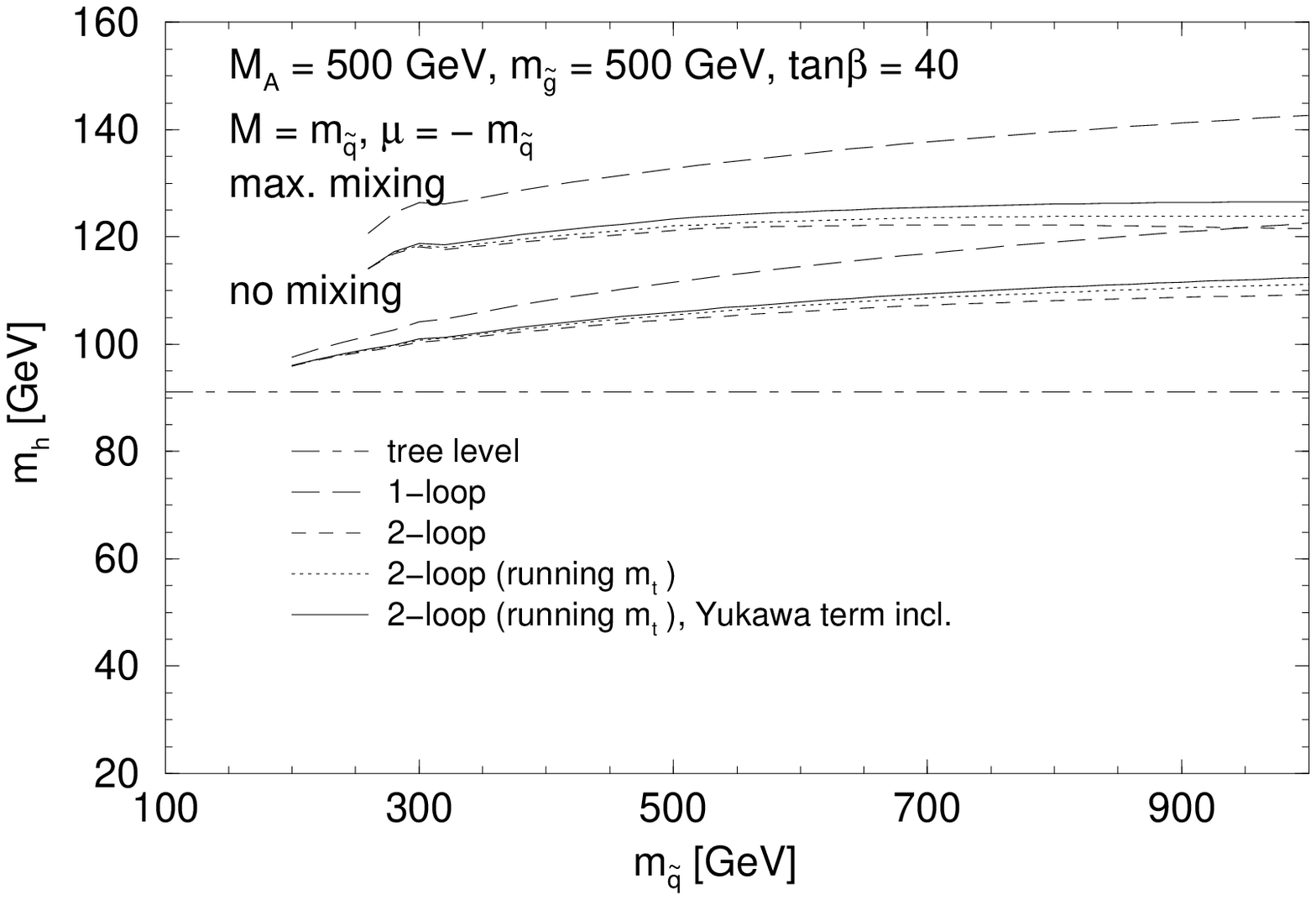,width=5.3cm,height=8cm,
                      bbllx=150pt,bblly=100pt,bburx=450pt,bbury=420pt}}
\end{center}
\caption[]{
The mass of the lightest Higgs boson for $\Tb = 1.6$ and $\Tb = 40$. 
The tree-level, the \onel\ and the \twol\ results for $\mh$ are shown 
as a function of $\msq$ for the no-mixing and the maximal-mixing case.
} 
\label{fig:plot2}
\end{figure}

In \reffi{fig:plot2} $\mh$ is shown in the two scenarios with 
$\Tb = 1.6$ and $\Tb = 40$ as a function of $\msq$ for no mixing and 
maximal mixing. The tree-level, the \onel\ and the \twol\ results with
the two steps of refinement are shown (the values of $\msq$ are such
that the corresponding $\Stop$-masses lie within the experimentally
allowed region). In all scenarios of
\reffi{fig:plot2} the \twol\ corrections give rise to a large reduction
of the \onel\ value of $\mh$. The biggest effect occurs for the $\Tb =
1.6$ scenario with maximal mixing. The inclusion of the refinement
terms leads to a slight shift in $\mh$ towards higher values, whose 
size is about $20\%$ of the \twol\ correction. In the $\Tb = 1.6$
scenario, $\mh$ reaches about $80 \gev$ for $\msq = 1 \tev$ in the 
no-mixing case, and about $100 \gev$ for $\msq = 1 \tev$ in the 
maximal-mixing case. For $\Tb = 40$ the respective values of $\mh$ are
nearly $115 \gev$ in the no-mixing case, and almost $130 \gev$ in the
maximal-mixing case.

%%%%%%%%%%%%%%%%%%%%%%%%%%%%%%%%%%%%%%%%%%%%%%%%%%%%%%%%%%%%%%
%%%%%%%%%%%%%%%%%%%%%%%%%%%%%%%%%%%%%%%%%%%%%%%%%%%%%%%%%%%%%%

Varying the gaugino parameter $M$, which enters via the non-leading \onel\ 
contributions, in the results of \reffi{fig:plot2} changes the value of
$\mh$ within $3 \gev$. Different values of the gluino mass, $\mgl$,
in the \twol\ contribution affect the prediction for $\mh$ by up to
$3 \gev$. Allowing for a splitting between the parameters 
$\MstL$, $\MstR$ in the $\Stop$ mass matrix yields maximal values of
$\mh$ which are approximately the same as for the case $\msq = \MstL =
\MstR$ (see also \reffi{fig:plot5} below).
Varying $\Tb$ around the value $\Tb = 1.6$ leads to a
relatively large effect in $\mh$ (higher values for $\mh$ are obtained
for larger $\Tb$), while the effect of varying $\Tb$ around $\Tb = 40$
is marginal. A more detailed analysis of the dependence of our results on the 
different SUSY parameters will be presented in a forthcoming
publication.
The discovery limit of LEP2 is expected to be slightly above 
$100 \gev$~\cite{lep2discpot}. Accordingly, our results confirm that
for the scenario with $\Tb = 1.6$ practically the whole parameter space
of the MSSM can be covered at LEP2.
For slightly larger $\Tb$ and maximal mixing, however, some parameter
space could remain in which the Higgs boson escapes detection at LEP2.
For $\Tb = 40$, on the other hand, a full exploration of the MSSM
parameter space will not be possible at LEP2. While the prediction for
$\mh$ is at the edge of the LEP2 range in the no-mixing case, the case
of large mixing will be a challenge for the upgraded Tevatron or will
finally be probed at LHC.

%%%%%%%%%%%%%%%%%%%%%%%%%%%%%%%%%%%%%%%%%%%%%%%%%%%%%%%%%%%%%%%%%%%%%%%
%%%%%%%%%%%%%%%%%%%%%%%%%%%%%%%%%%%%%%%%%%%%%%%%%%%%%%%%%%%%%%%%%%%%%%%

%%%%%%%%%%%%%%%%%%%%%%%%%%%%%%%%%%%%%%%%%%%%%%%%%%%%%%%%%%%%%%
\begin{figure}[htb]
\begin{center}
%\hspace{1em}
\mbox{
\psfig{figure=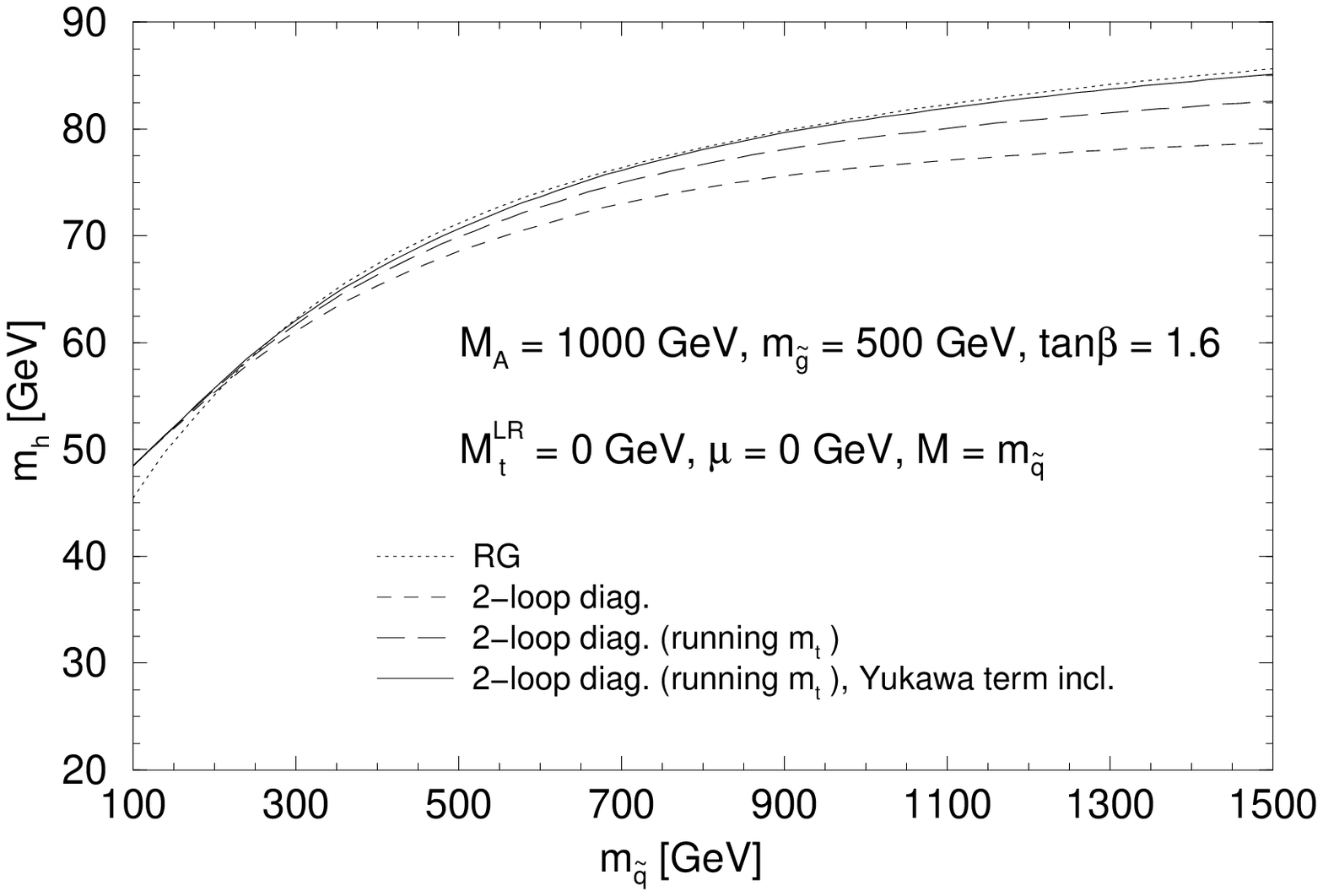,width=5.3cm,height=8cm,
                      bbllx=150pt,bblly=100pt,bburx=450pt,bbury=420pt}}
\hspace{7.5em}
\mbox{
\psfig{figure=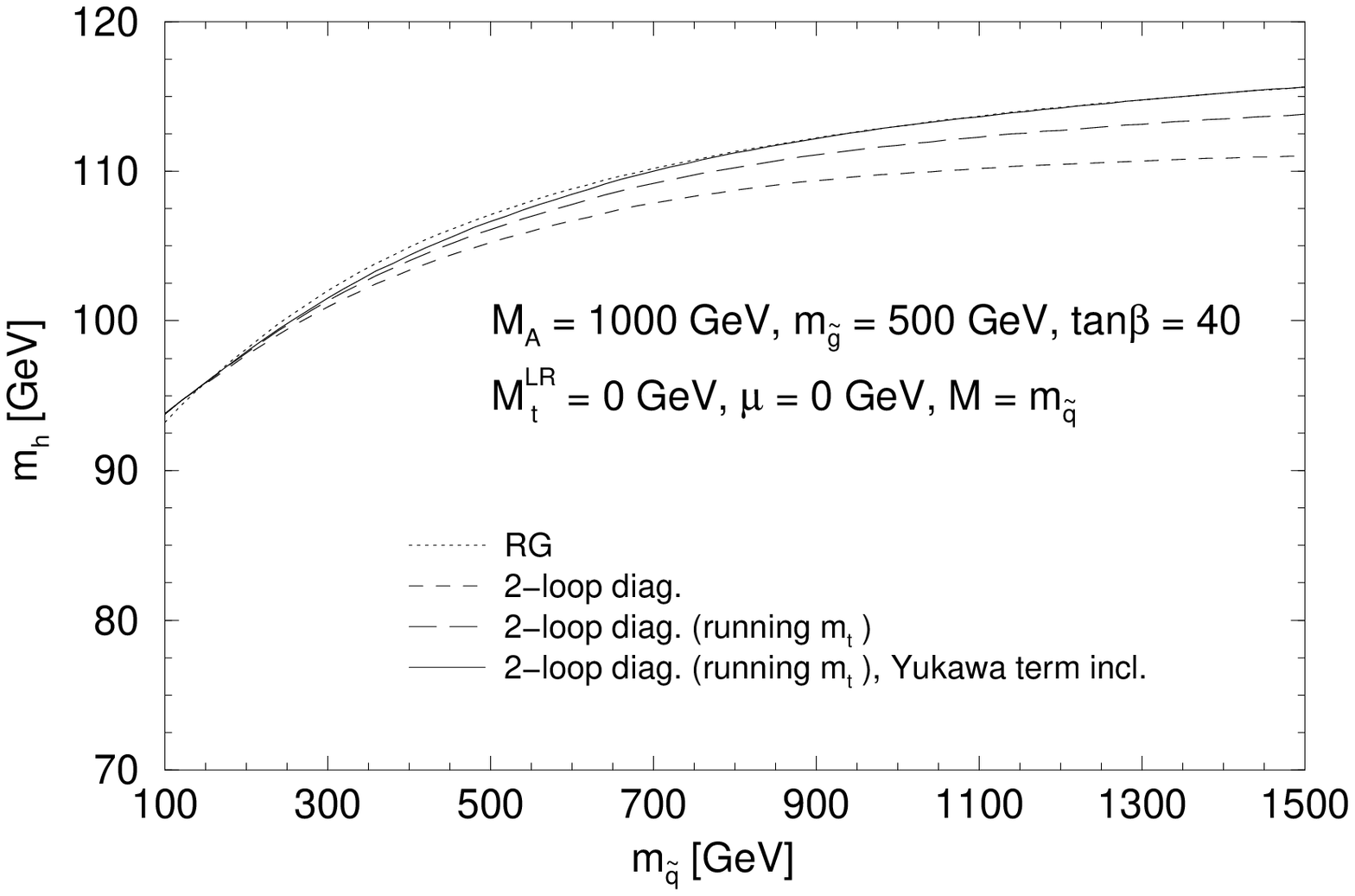,width=5.3cm,height=8cm,
                      bbllx=150pt,bblly=100pt,bburx=450pt,bbury=420pt}}
\end{center}
\caption[]{
Comparison between the Feynman-diagrammatic calculations and the
results obtained by renormalization group methods~\cite{mhiggsRG1b}.
The mass of the lightest Higgs boson is shown for the two scenarios with
$\Tb = 1.6$ and $\Tb = 40$ for the case of vanishing mixing in the
$\Stop$-sector.} 
\label{fig:plot3}
\end{figure}
%%%%%%%%%%%%%%%%%%%%%%%%%%%%%%%%%%%%%%%%%%%%%%%%%%%%%%%%%%%%%%

%%%%%%%%%%%%%%%%%%%%%%%%%%%%%%%%%%%%%%%%%%%%%%%%%%%%%%%%%%%%%%
\begin{figure}[bth]
\begin{center}
%\vspace{-5em}
%\hspace{1em}
\mbox{
\psfig{figure=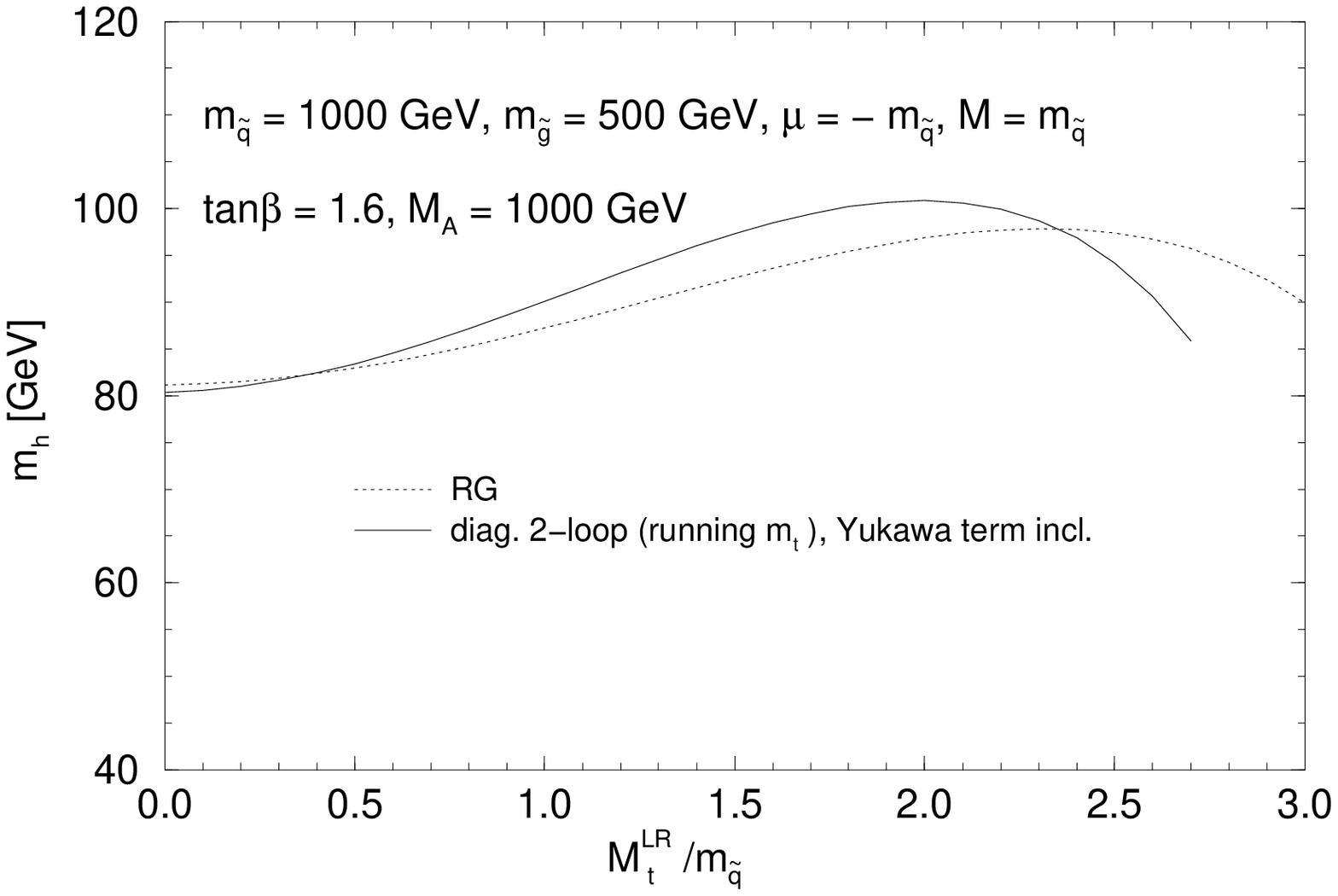,width=5.3cm,height=8cm,
                      bbllx=150pt,bblly=100pt,bburx=450pt,bbury=420pt}}
\hspace{7.5em}
\mbox{
\psfig{figure=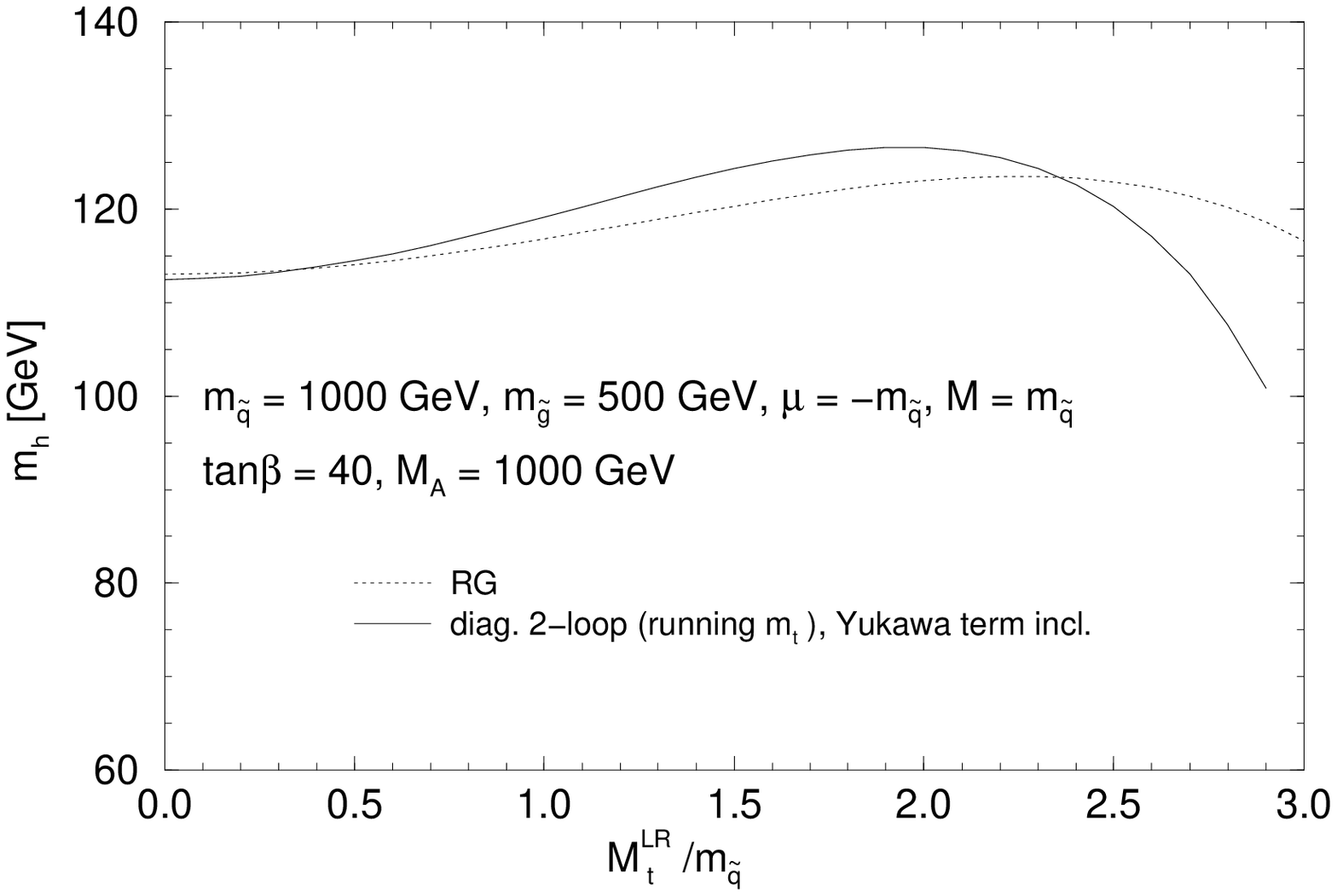,width=5.3cm,height=8cm,
                      bbllx=150pt,bblly=100pt,bburx=450pt,bbury=420pt}}
\end{center}
\caption[]{
Comparison between the Feynman-diagrammatic calculations and the
results obtained by renormalization group methods~\cite{mhiggsRG1b}.
The mass of the lightest Higgs boson is shown for the two scenarios with
$\Tb = 1.6$ and $\Tb = 40$ for increasing mixing in the
$\Stop$-sector and $\msq = \MA$.} 
\label{fig:plot4}
\end{figure}
%%%%%%%%%%%%%%%%%%%%%%%%%%%%%%%%%%%%%%%%%%%%%%%%%%%%%%%%%%%%%%

We now turn to the comparison of our diagrammatic results with the
predictions obtained via renormalization group methods. We begin with
the case of vanishing mixing in the $\Stop$~sector and large values 
of $\MA$, for which the RG approach is most easily applicable and is
expected to work most accurately. In order to study different
contributions separately, we have first compared the diagrammatic
one-loop on-shell result~\cite{mhiggsf1l} with the \onel\ leading log
result (without renormalization group improvement) given in
\citere{mhiggsRG2} and found very good agreement, typically within 
$1 \gev$. We then performed a leading log expansion of our diagrammatic
result
(which corresponds to the \twol\ contribution in the RG approach) and
also found agreement with the full \twol\ result
within about $1 \gev$. Finally, as shown in 
\reffi{fig:plot3}, we have compared our diagrammatic result for the 
no-mixing case including the refinement terms with the RG results
obtained in \citere{mhiggsRG1b}.%
\footnote{The RG results of \citere{mhiggsRG1b} and \citere{mhiggsRG2} 
agree within about $2 \gev$ with each other.}
As can be seen in \reffi{fig:plot3}, after the inclusion of the refinement 
terms the diagrammatic result for the no-mixing case agrees very well 
with the RG result. The deviation between the results exceeds
$2 \gev$ only for $\Tb = 1.6$ and $\msq < 150 \gev$.
For smaller values of $\MA$ the comparison for the no-mixing case
looks qualitatively the same. For $\Tb = 1.6$ and values of $\MA$ below 
$100 \gev$ slightly larger deviations are possible. Since the RG
results do not contain the gluino mass as a parameter, varying $\mgl$
gives rise to an extra deviation, which in the no-mixing case does not
exceed $1 \gev$. Varying the other parameters $\mu$ and $M$ in general
does not
lead to a sizable effect in the comparison with the corresponding RG
results.

We now consider the situation when mixing in the $\Stop$~sector is
taken into account. We have again compared the full \onel\ result with
the \onel\ leading log result used within the RG
approach~\cite{mhiggsRG2} and found good agreement. Only for values of
$\MA$ below $100 \gev$ and large mixing deviations of about $5 \gev$
occur. In \reffi{fig:plot4} our diagrammatic result including the
refinement terms is compared with the RG results~\cite{mhiggsRG1b}
as a function of
$\Mtlr/\msq$ for $\Tb = 1.6$ and $\Tb = 40$. The point $\Mtlr/\msq = 0$
corresponds to the plots shown in \reffi{fig:plot3}, except 
that the parameter $\mu$ is now set to $\mu = -\msq$. For larger
$\Stop$-mixing sizable deviations between the diagrammatic and the RG
results occur, which can exceed $5 \gev$ for moderate mixing and become
very large for large values of $\Mtlr/\msq$. 
As already stressed above, the maximal
value for $\mh$ in the diagrammatic approach is reached for
$\Mtlr/\msq \approx 2$, whereas the RG results have a maximum at 
$\Mtlr/\msq \approx 2.4$, i.e. at the \onel\ value. Varying the value
of $\mgl$ in our result leads to a larger effect than in the 
no-mixing case and shifts the diagrammatic result relative to the RG
result within $\pm 2 \gev$.

The results of our diagrammatic on-shell
calculation and the RG methods have been compared above in terms of the
parameters $\MstL$, $\MstR$ and $\Mtlr$ of the $\Stop$~mixing matrix,
since the available numerical codes for the RG 
results~\cite{mhiggsRG1b,mhiggsRG2} are given in terms of these
parameters. However, since the two approaches rely on different
renormalization schemes, the meaning of these (non-observable)
parameters is not precisely the same in the two approaches
starting from \twol\ order. 
Indeed we have checked that
assuming fixed values for the physical parameters $\mste$,
$\mstz$, and $\tst$ and deriving the corresponding values of the 
parameters $\MstL$, $\MstR$ and $\Mtlr$ in the on-shell scheme as well
as in the $\overline{MS}$ scheme, sizable differences occur between the
values of the mixing parameter $\Mtlr$ in the two schemes, while the 
parameters $\MstL$, $\MstR$ are approximately equal in the two schemes.
Thus, part of the different shape of the curves in \reffi{fig:plot4}
may be attributed to a different meaning of the parameter $\Mtlr$ in
the on-shell scheme and in the RG calculation.

%%%%%%%%%%%%%%%%%%%%%%%%%%%%%%%%%%%%%%%%%%%%%%%%%%%%%%%%%%%%%%
\begin{figure}[htb]
\begin{center}
\hspace{1em}
\mbox{
\psfig{figure=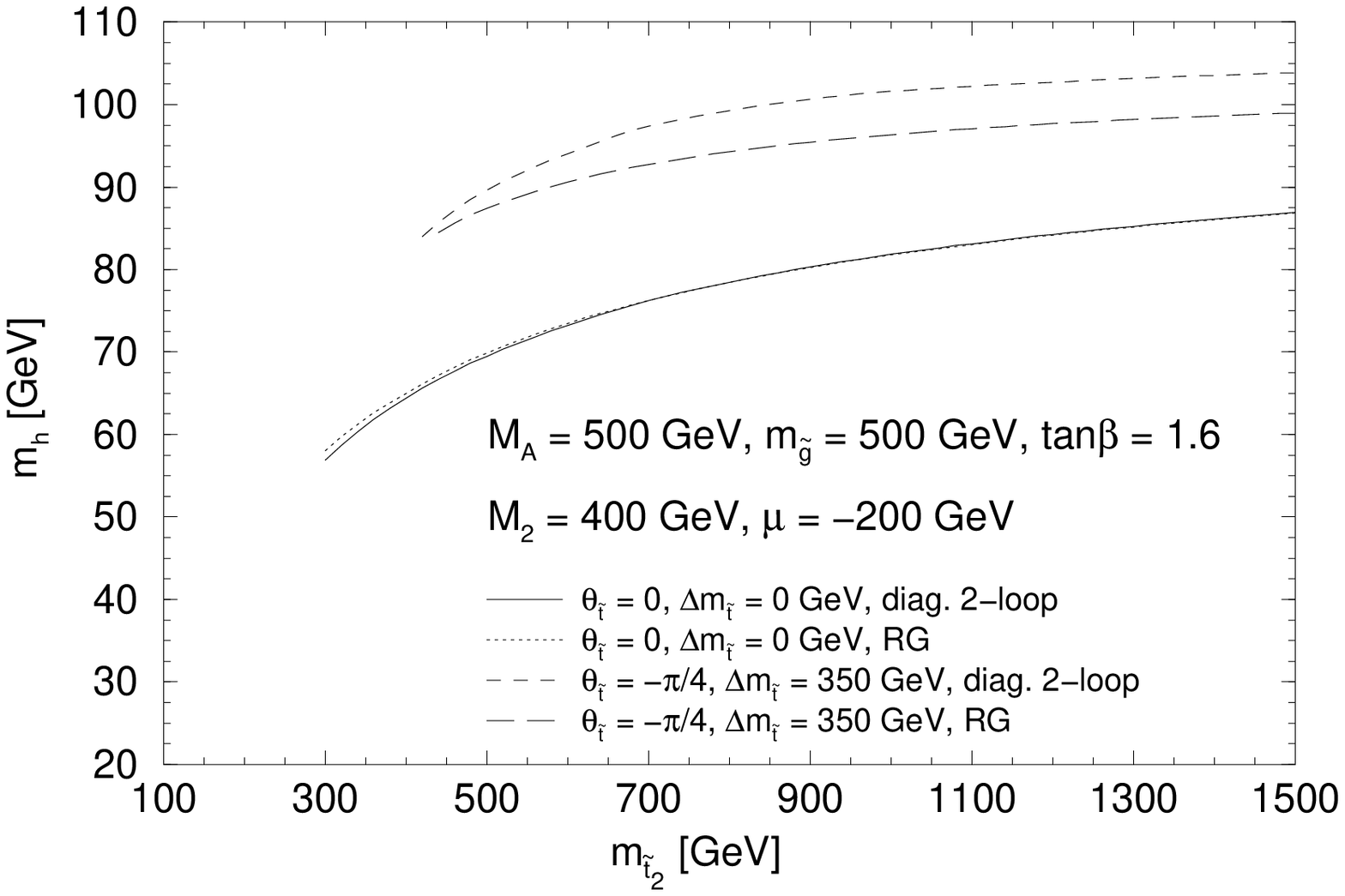,width=5.3cm,height=8cm,
                      bbllx=150pt,bblly=100pt,bburx=450pt,bbury=420pt}}
\hspace{7.5em}
\mbox{
\psfig{figure=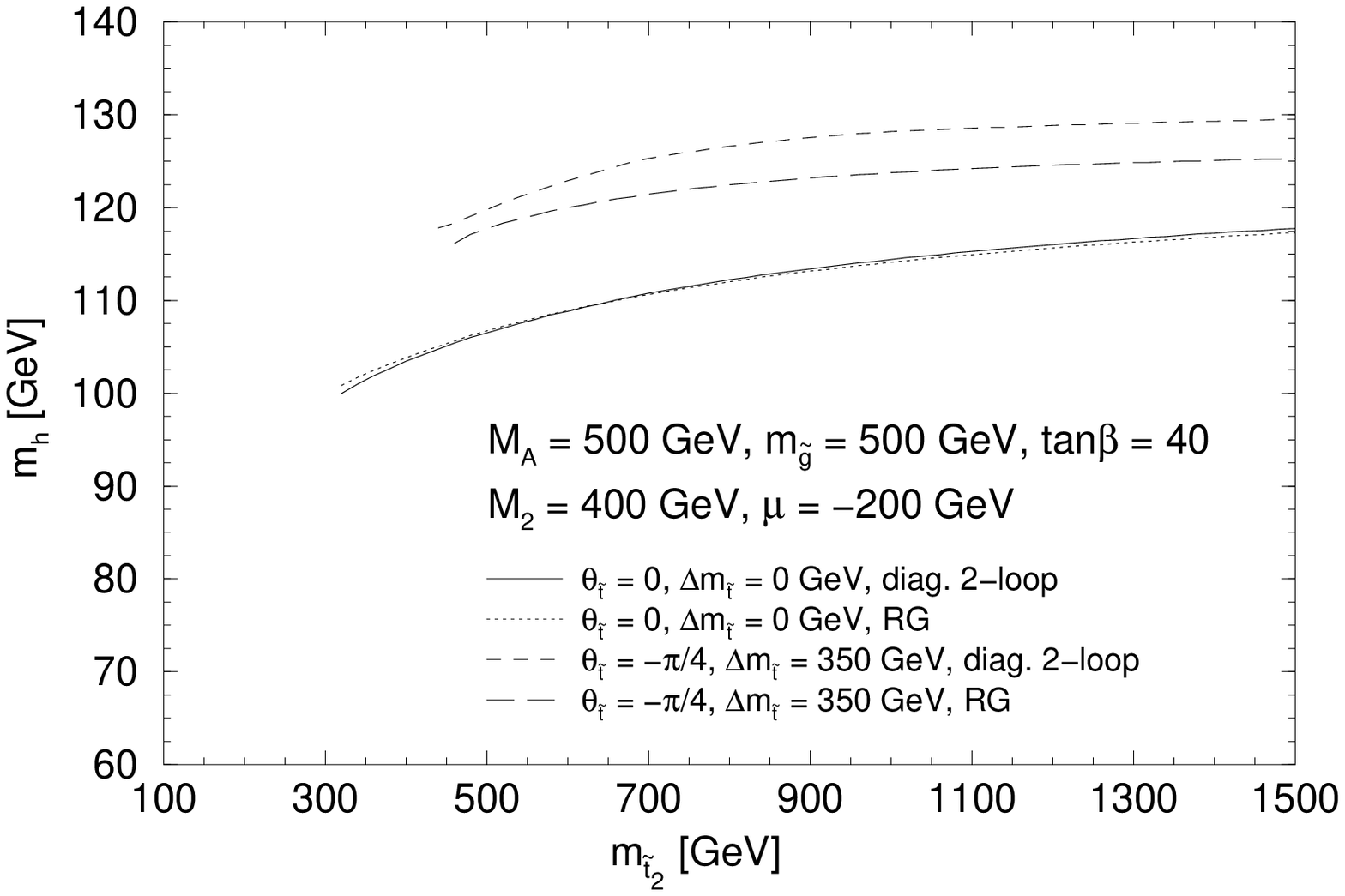,width=5.3cm,height=8cm,
                      bbllx=150pt,bblly=100pt,bburx=450pt,bbury=420pt}}
\end{center}
\caption[]{
Comparison between the Feynman-diagrammatic calculations and the
results obtained by renormalization group methods~\cite{mhiggsRG1b}.
The mass of the lightest Higgs boson is shown for the two scenarios with
$\Tb = 1.6$ and $\Tb = 40$ as a function of the heavier physical 
$\Stop$ mass $\mstz$.
For the curves with $\tst = 0$ a mass difference $\delmst = 0 \gev$ is
assumed whereas for $\tst = -\pi/4$ we chose $\delmst = 350 \gev$, for
which the maximal Higgs masses are achieved.}
\label{fig:plot5}
\end{figure}
%%%%%%%%%%%%%%%%%%%%%%%%%%%%%%%%%%%%%%%%%%%%%%%%%%%%%%%%%%%%%%

In order to compare results obtained by different approaches making use
of different renormalization schemes we find it preferable to compare
predictions for physical observables in terms of other observables
(instead of unphysical parameters). As a step into this direction we
compare in \reffi{fig:plot5} the diagrammatic results and the RG
results as a function of the physical mass $\mstz$ and with the mass
difference $\delmst = \mstz - \mste$ and the mixing angle $\tst$ as
parameters. 
In the context of the RG approach the running $\Stop$ masses, derived
from the $\Stop$ mass matrix, are considered as an approximation for
the physical masses.
The range of the $\Stop$ masses appearing in \reffi{fig:plot5} has been
constrained by requiring that the contribution of the third generation
of scalar quarks to the $\rho$-parameter~\cite{drhosuqcd} does not exceed 
the value of $1.3 \cdot 10^{-3}$, which corresponds to the resolution
of $\De\rho \;( = \epsilon_1 )$ when it is determined from experimental
data~\cite{delrhoexp}. 
As in the comparison 
performed above, in \reffi{fig:plot5} very good agreement is found between
the results of the two approaches in the case of vanishing
$\Stop$~mixing. For the maximal mixing angle $\tst = -\pi/4$, however,
the diagrammatic result yields values for $\mh$ which are higher by
about $5 \gev$.
\bigskip

In summary, we have implemented the result of the Feynman-diagrammatic
calculation of the leading $\oaas$ corrections to the masses of the
neutral $\cp$-even Higgs bosons in the MSSM into the prediction based
on the complete diagrammatic \onel\ on-shell result. Two further
refinements have 
been added in order to incorporate leading electroweak \twol\ and
higher-order QCD contributions. In this way we provide the at present
most precise prediction for $\mh$ based on Feynman-diagrammatic
results. The results are valid for arbitrary values of the relevant
MSSM parameters. The \twol\ corrections lead to a large reduction of
the one-loop on-shell result. Concerning the discovery and exclusion
potential of LEP2, our results confirm that the scenario with low 
$\Tb$ ($\Tb = 1.6$) in the MSSM should be covered at
LEP2, while the scenario with high $\Tb$ ($\Tb = 40$) is only
(partly) accessible for vanishing mixing in the scalar top sector. 
We have compared our results with the results obtained via
RG methods, and have analyzed in this context the 
\onel\ and \twol\ contributions separately. We have found that the
drastic deviations present between the \onel\ on-shell result and the
RG result are largely reduced. For the case of vanishing mixing in the
scalar top sector the diagrammatic and the RG results agree very well,
while sizable deviations exceeding $5 \gev$ occur when $\Stop$~mixing 
is taken into account. Since the gluino mass does not appear as a
parameter in the RG results, its variation gives rise to an extra
deviation which lies within $\pm 2 \gev$.
%in the case of large $\Stop$~mixing.
We have discussed the issue of how 
%It is very important to allow 
the results obtained via different approaches using 
different renormalization schemes can be formulated such that they are
readily comparable to each other
also when corrections beyond \onel\ order are incorporated.  
For this purpose it is very desirable to express 
the prediction for the
Higgs-boson masses in terms of physical observables, i.e.\ the physical 
masses and mixing angles of the model.
\bigskip

We thank M.~Carena, H.~Haber and C.~Wagner for fruitful discussions and 
communication about the numerical comparison of our results. We also 
thank A.~Djouadi and M.~Spira for helpful discussions.

%%%%%%%%%%%%%%%%%%%%%%%%%%%%%%%%%%%%%%%%%%%%%%%%%%%%%%%%%%%%%%
%%%%%%%%%%%%%%%%%%%%%%%%%%%%%%%%%%%%%%%%%%%%%%%%%%%%%%%%%%%%%%

%\newpage

%%%%%%%%%%%%%%%%%%%%%%%%%%%%%%%%%%%%%%%%%%%%%%%%%%%%%%%%%%%%%%
%%%%%%%%%%%%%%%%%%%%%%%%%%%%%%%%%%%%%%%%%%%%%%%%%%%%%%%%%%%%%%

\end{document}